\documentclass[amsmath,amssymb, aps, prl, twocolumn]{revtex4-1}

\usepackage{graphicx}
\usepackage{dcolumn}
\usepackage{bm}
\usepackage[usenames]{color}
\begin{document}

\title{Quantum Loop Topography for Machine Learning
}

\author{Yi Zhang}
\email{frankzhangyi@gmail.com}
\author{Eun-Ah Kim}
\email{eun-ah.kim@cornell.edu}

\affiliation{Department of Physics, Cornell University, Ithaca, New
York 14853, USA} \affiliation{Kavli Institute for Theoretical
Physics, University of California, Santa Barbara, California 93106,
USA}

\date{\today}

\begin{abstract}
Despite rapidly growing interest in harnessing machine learning in the study of quantum many-body systems, training neural networks to identify quantum phases is a nontrivial challenge. The key challenge is in efficiently extracting essential information from the many-body Hamiltonian or wave function and turning the information into an image that can be fed into a neural network. When targeting topological phases, this task becomes particularly challenging as topological phases are defined in terms of non-local properties. Here we introduce quantum loop topography (QLT): a procedure of constructing a multi-dimensional image from the ``sample'' Hamiltonian or wave function by evaluating two-point operators that form loops at independent Monte Carlo steps. The loop configuration is guided by characteristic response for defining the phase, which is Hall conductivity for the cases at hand. Feeding QLT to a fully-connected neural network with a single hidden layer, we demonstrate that the architecture can be effectively trained to distinguish Chern insulator and fractional Chern insulator from trivial insulators with high fidelity. In addition to establishing the first case of obtaining a phase diagram with topological quantum phase transition with machine learning, the perspective of bridging traditional condensed matter theory with machine learning will be broadly valuable.
\end{abstract}

\maketitle

{\it Introduction--}
Machine learning techniques have been enabling neural networks to  successfully recognize and interpret big data sets of images and speeches\cite{MLbook}. Through supervised trainings with a large number of data sets, neural networks `learn' to recognize key features of a universal class. Very recently, rapid and promising development has been made from this perspective on numerical studies of condensed matter systems, including dynamical systems\cite{Schoenholz2016, BM1, BM2, BM3, BM4}, systems undergoing phase transitions\cite{Melko20161, LeiWang2016, Simon2016, Kelvin2016, Meng2016, Wanglei2016, Ohtsuki2016}, as well as quantum many-body systems. Also established is the theory connection to renormalization group\cite{Beny2013, Schwab2014}. Exciting successes in application of machine learning to symmetry broken phases\cite{Melko20161, LeiWang2016, Simon2016, Kelvin2016} may be attributed to the locality of the defining property of the target phases: the order parameter field. The snap-shots of order parameter configuration form images that can be readily fed into neural networks that have been developed to recognize patterns in images.

Unfortunately many novel states cannot be numerically detected through a local order parameter. For one, all topological phases are intrinsically defined in terms of non-local topological properties. Not only many-body localized states of growing interest\cite{MBLreview} fit into this category, even a superconducting state fits in here since the superconducting order parameter explicitly breaks particle number conservation\cite{Scalapino1993}. In order for neural networks to learn to recognize and identify such phases, we need to supply them with ``images'' that contain relevant non-local information. Clearly information based on single site is insufficient. One approach to detecting topological phase was to augment single site based information with additional layers of convolutional filters that add complexity to the neural network architecture and implementing local constraints relying on translational symmetry, targeting a single topological phase at a time\cite{Melko20161, Kelvin2016}. Another approach was to detect the topological phase's edge states\cite{Ohtsuki2016}. In addition, ensemble of the Green's function was used to detect charge-ordered phases\cite{Simon2016}.

Here we introduce quantum loop topography (QLT): a procedure that designs and selects the input data based on the target phases of interest guided by relevant response functions. We focus on the fermionic topological phases but the procedure can be generalized to other situations that are not captured by purely local information, as all physically meaningful states are characterized by their response functions. The subject of topological phases of matter has grown with the appeal that topological properties are non-local and hence more robust\cite{Kitaev20032, Chetan2005, Chetan2008}. Ironically this attractive feature makes it difficult to detect and identify topological phases even in numerics. Importantly, detection of strongly-correlated topological phases as fractional quantum Hall states\cite{LaughlinFQH, TsuiFQH}, fractional Chern insulators\cite{Titus2011, BernevigFCI}, quantum spin liquids\cite{Laughlin1987, Wen1990, Balents2010} requires arduous calculations of topological entanglements entropies\cite{KitaevTee, WenTee}. On the other hand, quantization\cite{IQHE1980,Ludwig1994,Haldane1988,TI2010,TKNN1982,LaughlinFQH, TsuiFQH,Titus2011, BernevigFCI} is a natural theme of all topological states and one may wonder perhaps there can be an intelligent way to detect topological phases due to the discreteness in defining properties. In this letter we demonstrate that QLT enables even a rather simple neural network architecture consisting of a fully-connected neural network with a single hidden layer to recognize Chern insulator and fractional Chern insulator states and rapidly produce a phase diagram containing topological quantum phase transition. We then discuss insights into the effectiveness of QLT and future directions based on its versatility.

{\it Quantum Loop Topography and our algorithm--} The procedure we dubbed QLT constructs an input image from a given Hamiltonian or many-body wave function
that contains minimal but sufficient amount of non-local information guided by relevant response functions. The response function that characterizes the phase of interest determines the geometry of the loop objects that enter QLT. But instead of brute force evaluation of the response functions, we use QLT obtained from instances of Monte Carlo steps to train a network deep in the phases.

\begin{figure}
\includegraphics[scale=0.35]{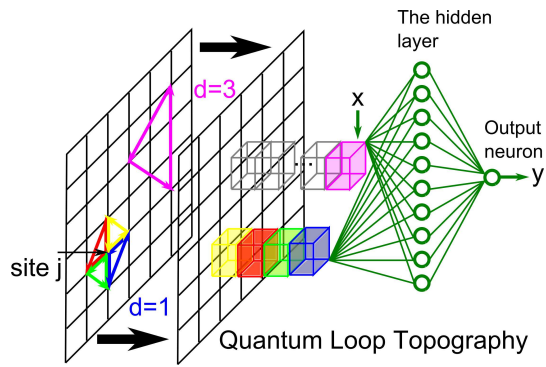}
\caption{Schematic illustration of our machine learning algorithm consisting of QLT and a neural network architecture. QLT for each site $j$ consists of 4 loops of length $d=1$. One loop of length $d=3$ is also shown for illustration. QLT of length $d \leq d_c$ form a $D(d_c)$-dimensional vector for each site $j$, e.g., $D(1)=4$ on a square lattice. }
\label{fig:machine}
\end{figure}

For Chern insulators of interest here, the relevant response function is the Hall conductivity. Interestingly \textcite{Kitaev20062} pointed out that
\begin{equation} \sigma_{xy} = \frac{e^{2}}{h}\cdot\frac{1}{N} \sum 4\pi iP_{jk}P_{kl}P_{lj}S_{\triangle jkl}
\label{eq:sigmaxy}
\end{equation}
for free fermion systems\footnote{Our alternative proof of
Eq.~\eqref{eq:sigmaxy} (see Supplemental Material) builds on
adiabatic continuity protected by the gap in the spectra without
requiring the system be non-interacting.}, where
$P_{ij}\equiv\langle c_i^\dagger c_j\rangle$ is the equal-time
two-point correlation functions between site $i$ and site $j$,
$S_{\triangle jkl}$ is the signed area of the triangle $jkl$, and
$N$ is the total number of sites. Taking hints from
Eq.~\eqref{eq:sigmaxy} we use triangular loops to define QLT for
Chern insulators. But instead of the full expectation value for
two-point correlation functions in Eq.~\eqref{eq:sigmaxy} which are
costly to evaluate (requiring many instances of Monte Carlo walking
down the Markov chain), we evaluate the bilinear operator with a
single Monte Carlo sample defining  $\tilde{P}_{jk}|_{\alpha}\equiv
\left\langle c^\dagger_j c_k \right\rangle_\alpha$ for  a particular
Monte Carlo sample $\alpha$. Further we note that smaller triangles
will dominantly contribute in a gapped system and keep the loops of
linear dimension less than a cut-off $d_c$.

Now we define QLT to be a quasi-two-dimensional ``image" of $D(d_c)$-dimensional vector of complex numbers assigned to each lattice site $j$, where $d_c$ is the cut-off length and $D(d_c)$ is the total number of triangles of length $d\leq d_c$ with one vertex at site $j$ (see Fig. \ref{fig:machine}). Each entry of this vector is associated with a distinct triangle cornered at site $j$ which defines a chained product
\begin{equation}
\tilde{P}_{jk}|_{\alpha}\tilde{P}_{kl}|_{\beta}\tilde{P}_{lj}|_{\gamma}
\label{eq:tql}
\end{equation}
where $k$ and $l$ are two other sites of the particular triangle and
$\tilde{P}$'s are evaluated at three independent Monte Carlo steps
without averaging over Markov chain. This way, QLT can be
systematically expanded to include longer ranged correlations
involving site $j$ by increasing cut-off length scale $d_c$. When
the outcome converges for small $d_c$, QLT is quasi-two-dimensional.

By construction QLT is quite versatile. Firstly, QLT can be obtained for different lattice geometry to form a diverse input data as different lattice geometry only enter through different dimension $D(d_c)$ for given $d_c$. Secondly, the entire procedure takes place in real space  without any need for diagonalization or flux insertion and the procedure does not depend on translational invariance. Hence QLT should be able to naturally accommodate heterogeneity, disorder and interaction by construction. Finally, it is clear that the strategy underneath QLT construction for fermionic topological phases we have laid out here can be generalized for detection of other novel phases such as $\mathbb Z_2$ topological order or superconductivity\cite{mlz2topo}. In the rest of this paper we use Variational Monte Carlo(VMC), without loss of generality, to build QLT by sampling the many-body ground state of interest at randomly selected Monte Carlo steps (see Supplemental Material).

Once QLT is obtained for a given model, we feed it to a neural
network(Fig.~\ref{fig:machine}). For this, we designed a
feed-forward fully-connected neural network with only one hidden
layer consisting of $n=10$ sigmoid neurons. The network takes QLT as
an input $x$ and each neurons processes the input through
independent weights and biases $w\cdot x+b$. After the sigmoid
function, the outcome is fed forward to be processed by the output
neuron. The final output $y$ corresponds to the neural network's
judgement whether the input QLT is topological. We use cross entropy
as the cost function with L2 regularization to avoid over-training
and a mini-batch size of 10\cite{MLbook}. For the rest of this
paper, we use randomly-mixed 20000 data samples within the VMC
Metropolis of the topological and trivial phases as the training
group. We reserve a separate group of 4000 data samples (also half
trivial and half topological) for validation purposes including
learning speed control and termination\cite{MLbook}. Once the neural
network is successfully trained, the trained network can rapidly
process QLT's from different parts of the phase space to yield a
phase diagram. In order to establish level of confidence on the
trained network's assessment of whether the system is topological or
not, we process 2000 QLT's at each point and take the ratio $p$ of
`topological' output, i.e., $y>0.5$. When $p$ is close to 1 for
topological phase and 0 for trivial phase, it indicates even a
single QLT can reliably land a trustworthy detection.

\begin{figure}
\includegraphics[scale=0.35]{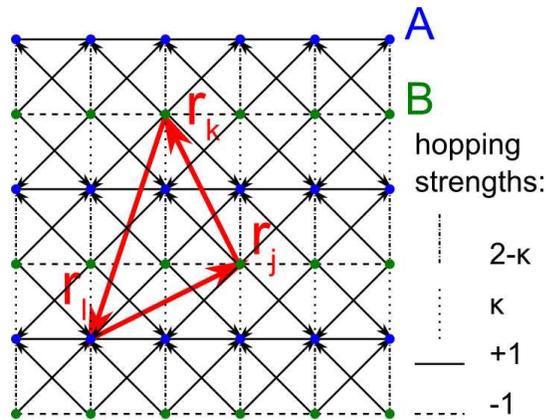}
\caption{Model illustration of Eq. \ref{eq:sqham}. The unit cell consists of two sublattice sites $A$ and $B$. Hopping strengths are different for horizontal and vertical bonds and
staggered. The diagonal hopping is $i\kappa$ ($-i\kappa$) along (against) the arrow. The red arrows denotes a triangle that defines the operators of our {\bf QLT}.}\label{fig:model}
\end{figure}

{\it Topological quantum phase transition in a free fermion model--} We first apply the QLT-based machine learning to the topological quantum phase transition between a trivial insulator and a Chern insulator. Consider the following tight-binding model on a square lattice:
\begin{eqnarray}
H(\kappa)&=&\underset{\vec{r}}{\sum}(-1)^{y} c_{\vec{r}+\hat x}^{\dagger}c_{\vec{r}} + [1+(-1)^{y} (1-\kappa)] c_{\vec{r}+\hat y}^{\dagger}c_{\vec{r}}\nonumber \\
&+& (-1)^{y}  \frac{i\kappa}{2}\left[c_{\vec{r}+\hat x + \hat y}^{\dagger}c_{\vec{r}}+c_{\vec{r}+\hat x - \hat y}^{\dagger}c_{\vec{r}}\right] + \mbox{h.c.}
\label{eq:sqham}
\end{eqnarray}
where $\vec{r} = (x,y)$ (see Fig.~\ref{fig:model}) and $\kappa$ is a  tuning parameter with $0\leq\kappa\leq1$. The $\kappa=1$ limit is the $\pi$-flux square lattice model for a Chern insulator with a Chern number $C=1$~\cite{Ludwig1994}, while the $\kappa=0$ limit amounts to decoupled two-leg ladders. $H(\kappa)$ interpolates between a Chern insulator and a trivial insulator with a topological quantum phase transition at $\kappa=0.5$. To observe the quantum phase transition, one should assume translational invariance and Fourier transform the Hamiltonian Eq.~\eqref{eq:sqham} to detect the change in the integral of the Berry curvature of the band structure
\begin{eqnarray}
H\left(\kappa\right)&=&\underset{k}{\sum} \left[ 2\cos k_{y}+2i\sin k_{y}\left(1-\kappa+\kappa\sin k_{x}\right)\right] c^\dagger_{k,A}c_{k,B} \nonumber\\& & +2\cos k_{x} (c^\dagger_{k,A}c_{k,A}-c^\dagger_{k,B}c_{k,B})+\mbox{h.c.}
\end{eqnarray}
where $A$ and $B$ label the two sublattices. For this simple two-band model with two Dirac points at $(\pi/2,\pi/2)$ and $(-\pi/2,\pi/2)$
the topological quantum phase transition can be predicted by simply noting the change of the sign of the Dirac masses across $\kappa=0.5$.

\begin{figure}
\begin{centering}
\includegraphics[scale=0.35]{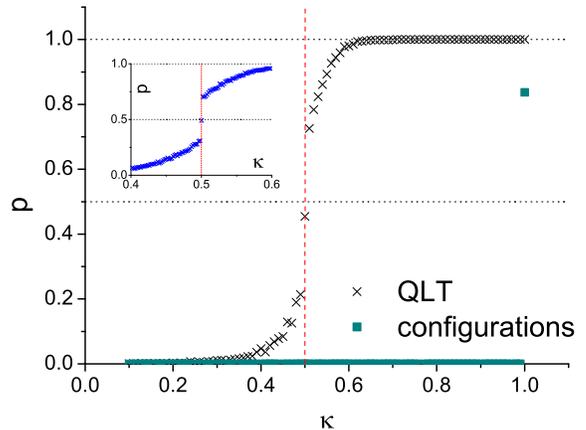}
\caption{The ratio $p$ of `topological' response from the neural network on the model in Eq. \ref{eq:sqham} over the parameter region $\kappa \in [0.1, 1,0]$. The neural network is trained with $\kappa=0.1$ for $y=0$ and $\kappa=1$ for $y=1$. The green square symbols represent the results using fermion occupation configurations as an input data. Red dashed line marks the expected topological phase transition at $\kappa=0.5$. The inset: an enlarged view over the critical region $0.4 \le\kappa\le 0.6$. $d_c=2$ for all.}
\label{fig:free}
\end{centering}
\end{figure}

Our complete knowledge of its topological phase diagram makes the model in Eq.~\ref{eq:sqham} an ideal testing ground for our algorithm. Hence we implement supervised machine learning on the models using two extreme points of $\kappa=1.0$ (Chern insulator) and $\kappa=0.1$ (trivial insulator) for training\footnote{The system becomes decoupled two-leg ladders at $\kappa=0.0$, which is non-generic for a two-dimensional insulator and we choose to avoid such specialty.}. The system size is $12\times 12$ lattice spacings unless noted otherwise.
First we establish that indeed a single point based input of the fermion occupation configurations $n(\vec{r})=c^{\dagger}_{\vec{r}}c_{\vec{r}}$ fails to transmit the topological information to the neural network, as we expected. With $n(\vec{r})$  as an input, the learning is inefficient and the neural network has difficulty picking up a clear structure even after a long period of training. Such struggle is signaled by high yields in the cost function\cite{MLbook}. Moreover, as shown in Fig.~\ref{fig:free}, the neural network keeps incorrectly judging the system to be a trivial insulator for all values of $\kappa$, except for $\kappa = 1.0$ where the result returns $>80\%$ `nontrivial'. This indicates that the neural network unfortunately does not pick up the universal features about the topological phase, but rather memorizes the more detailed information of the specific model at $\kappa=1.0$ itself.

The contrast in the results based on QLT input is striking. Fig.~\ref{fig:free} shows that the trained network's assessment achieves $>99.9\%$ accuracy deep in either the topological phase or trivial phase even with $d_c=2$. Moreover even though we have provided the training group with only large-gap models in both the topological and the trivial phases focusing on identifying phases\footnote{See Supplemental Material for further discussion and details on the impact of training models and QLT cut-off $d_c$ on machine learning phases as well as phase transitions.}, we find a non-analytical behavior in $p$ as a function of $\kappa$ at the critical point [see Fig. \ref{fig:free} inset]. Note the symmetric departure from $p\approx0.5$ on both sides of $\kappa=0.5$ reflects the symmetry in gap closing and reopening in the model of Eq.~\eqref{eq:sqham} which is not generic.

\begin{figure}[t]
\begin{centering}
\includegraphics[scale=0.35]{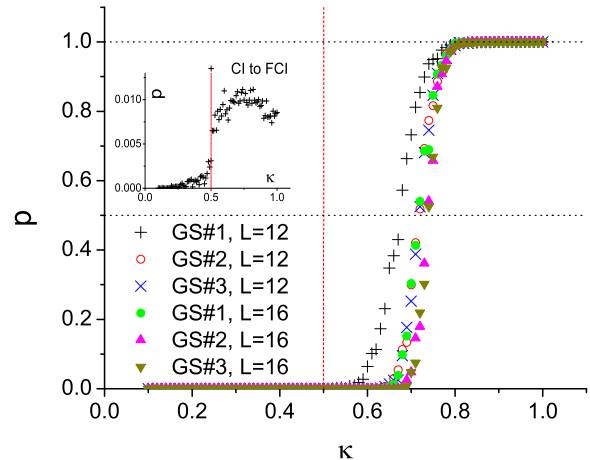}
\caption{Application to a $\nu=1/3$ FCI. The topological phase
transition in the {\it parent} Chern insulator at $\kappa =0.5$ is
marked by a vertical red dashed line. The inset shows the results
using neural network trained with the parent free fermion model,
where $p$ is calculated over 20000 samples for each $\kappa$ to
reduce statistical error. The main panel shows the results using FCI
wave functions for both training ($\kappa =0.1$ for trivial and
$\kappa =1.0$ for the FCI, first ground state only) and testing (all
three degenerate ground states, see Supplemental Material). $L=16$ data is shown in addition to $L=12$ to help attribute the differences between $\kappa_c$ of the topological phase transitions to the finite-size effect. $d_c=2$ for all. } \label{fig:fci}
\end{centering}
\end{figure}

{\it Generalizations--} Next we consider a fractional Chern insulator (FCI) as an example of strongly-correlated topological phase. Here the $\nu=1/3$ FCI is represented by a VMC wave function that is the free fermion wave function of the model in Eq. \ref{eq:sqham} raised to the third power\cite{FrankTEE}. Surprisingly the neural network trained on
non-interacting  parent Chern insulator already serves as a `poor man's network' (see the inset of Fig.~\ref{fig:fci}). This network recognizes that FCI phase is distinct from the parent Chern insulator and hence it only gives $p\sim 0.01$ `nontrivial' response for the FCI phase. Nevertheless it also notices that FCI is a topologically distinct state from the trivial insulator since $p\sim 0.01$ is large enough to exclude statistical error. Once trained with the FCI wave function at two reference points $\kappa=0.1$ for trivial and $\kappa=1.0$ for FCI, the network once again detects FCI phase with high accuracy.

Remarkably the network automatically recognizes topological degeneracy. Even when we train the network with one wave function deep in the trivial and topological phases (GS\#1 in Fig.~\ref{fig:fci} ), it correctly assess topological nature of two other wave functions that are related to the GS\#1 by flux threading. Moreover the network detect topological quantum phase transition at $0.67\le\kappa_c\le0.77$. The uncertainty in the critical value $\kappa_c$ is a finite-size effect as it is clear from the fact that degenerate wave functions converge to the same transition point upon increasing the system size [see Fig.~\ref{fig:fci}]. The fact that $\kappa_c>0.5$ when the single particle gap closes at $\kappa=0.5$ could raise concern in light of the findings on single particle Green's function based approaches\cite{Essin2013,Meng20162}. Nevertheless it is to be expected single particle gap is a pre-requisite for the VMC wave function to represent a topological phase since only then partons may be integrated out. Hence if anything, the shift of $\kappa_c>0.5$ is consistent with the expectations from the parton construction. Nevertheless, the result calls for further study for locating the critical point using an independent measure such as many-body gap. However, it is important to note that this is the first report of the topological quantum phase transition providing the target, which would have been too time-consuming with the more established topological entanglement entropy based approaches\cite{KitaevTee, WenTee, FrankTEE, smat, smat2}.

\begin{figure}[t]
\begin{centering}
\includegraphics[scale=0.35]{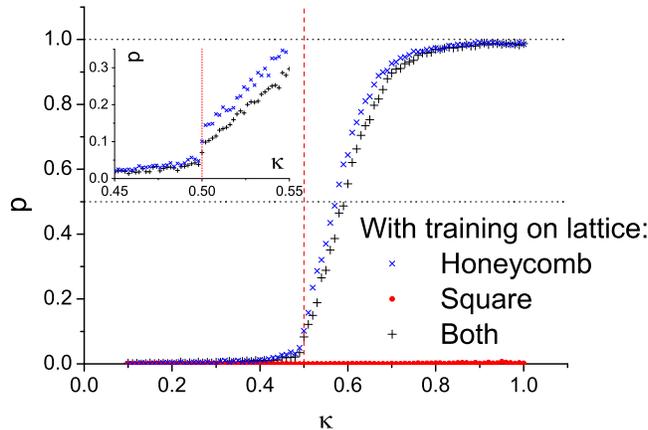}
\caption{The ratio $p$ of `topologically nontrivial' response from the neural networks for the honeycomb lattice model (Supplemental Material) over the parameter region $\kappa \in [0.1, 1,0]$. The topological phase transition is at $\kappa =0.5$ (vertical red dashed line). The neural networks are trained using the Chern insulators and trivial insulators only on the honeycomb lattice, only on the square lattice, and on both. The inset zooms into the critical region $0.4\le\kappa\le0.6$. $d_c=2$ for all.}
\label{fig:general}
\end{centering}
\end{figure}

Finally, we demonstrate that we can train the network to learn the
topological protection of topological order. The topological
protection implies indifference to the microscopic details such as
lattice structure or impurities. The key to a successful training on
this celebrated feature is the diversity of the training input.
Without diverse input, the network looks for features that are
specific to its training set. For instance, the network trained only
with square lattice cannot recognize the topological phase in the
honeycomb lattice. But if we provide diverse input taken from both
the square lattice and the honeycomb lattice systems, the network
can be trained to recognize topological phases on both lattices with
little penalty on accuracy (see Fig.~\ref{fig:general}). We also
note that the network recognizes the difference between different
Chern numbers
 (e.g., $C=-1$ v.s.  $C=1$).

{\it Conclusion--} In summary, we have successfully implemented supervised machine learning for topological phases by introducing QLT as an interface between traditional concept of response theory and a simple neural network.

Three major strengths of our QLT-based machine learning approaches are 1) efficiency, 2) accuracy, and 3) versatility. Firstly, the network can be trained with quasi-two dimensional QLT in gapped phases. Furthermore since QLT bypasses the time-consuming process of averaging over Markov chains, one can quickly scan the phase space once the network is trained. Although our focus was on the phases, we demonstrated that non-analyticity in the ratio of non-trivial response allows us to pinpoint the phase transition. Finally, as a real-space based formulation that does not requires translational symmetry, or diagonalization or flux insertion, QLT is quite versatile.

Specifically, our approach can be applied to systems with disorder or with higher Chern numbers as well as to higher dimensional systems. The fact that QLT readily handles degenerate ground states adds to its versatility. Moreover there is nothing restricting QLT to VMC data. It can be applied quantum Monte-Carlo samples of Hamiltonian based approaches \cite{Essin2013,Meng20162} as well as other representations of many-body wave functions such as matrix product states and PEPS. Most importantly, the procedure of defining appropriate QLT guided by relevant response function we established here for the specific case of topological phases is readily expanded to other state of interest such as superconducting state and $\mathbb Z_2$ topological order\cite{mlz2topo}. Hence our construction in this letter opens door to application of the machine learning approaches to novel states of broad interest.

\vspace{2mm}
\noindent {\bf Acknowledgements} We thank E. Khatami, R. Melko, T. Neupert and S. Trebst for insightful discussions. This work was supported by the DOE under Award DE-SC0010313. YZ acknowledges support through the Bethe Postdoctoral Fellowship and E-AK acknowledges Simons Fellow in Theoretical Physics Award \#392182. Bulk of this work was done at KITP supported by Grant No. NSF PHY11-25915.

\bibliographystyle{apsrev4-1}
\bibliography{refs}

\appendix

\section{Hall conductivity from two-point correlators}\label{app:sigmaxy}

In this section, we prove Eq. \ref{eq:sigmaxy}, that the Hall conductivity of a two-dimensional insulator can be expressed in terms of triangular loop-products of two-point
correlators $P_{ij}=\left\langle c_{i}^{\dagger}c_{j}\right\rangle $.
To start, we use the Kubo formula for Hall conductivity,
\begin{eqnarray*}
\sigma_{xy} & = & \frac{ie^{2}\hbar}{N}\left[\underset{n\neq0}{\sum}\frac{\left\langle \Phi_{0}\left|v_{y}\right|\Phi_{n}\right\rangle \left\langle \Phi_{n}\left|v_{x}\right|\Phi_{0}\right\rangle - x \leftrightarrow y  }{\left(E_{n}-E_{0}\right)^{2}}\right]\\
 & = & \frac{ie^{2}\hbar}{N}\left[\underset{m\in v}{\sum}\underset{n\notin v}{\sum}\frac{\left\langle m\left|v_{y}\right|n\right\rangle \left\langle n\left|v_{x}\right|m\right\rangle - x \leftrightarrow y  }{\epsilon_{n}^{2}}\right]
\end{eqnarray*}
where $\left|m\right>$ ($\left|n\right>$) are the single-particle states in the valence (conducting) bands.

On the other hand, the two-point correlators can be regarded as an operator that projects to the ground state $P=\underset{m\in
v}{\sum}\left|m\right\rangle \left\langle m\right|$. It also suggests that there exists a Hamiltonian $H'=-\Delta P$ with a flattened single-particle dispersion relation and an insulating gap of $\Delta$, which is adiabatically connected to the original system of interest (without closing the insulating gap).

It is straightforward to see that $\sigma_{xy}'$ can be further simplified with the replacement
$v_{x}'=\frac{i}{\hbar}\left[H',x\right]=-\frac{i\Delta}{\hbar}\left[P,x\right]$, $v_{y}'=\frac{i}{\hbar}\left[H',y\right]=-\frac{i\Delta}{\hbar}\left[P,y\right]$,
\begin{eqnarray*}
\sigma_{xy}' & = & \frac{ie^{2}\hbar}{N\Delta^2}\mbox{tr}\left[P v_y'\left(1-P\right)v_x'-P v_x' \left(1-P\right)v_y'\right]\\
 & = & -\frac{ie^{2}}{\hbar N}\mbox{tr}\left[P\left[P,y\right]\left[P,x\right]-P\left[P,x\right]\left[P,y\right]\right]\\
 & = & \frac{ie^{2}}{\hbar N} \sum  P_{jk}P_{kl}P_{lj}\left[ \left(\vec{r}_{k}-\vec{r}_{j}\right)\times\left(\vec{r}_{l}-\vec{r}_{j}\right)\cdot \hat z \right] \\
 & = & \frac{e^{2}}{h}\cdot\frac{1}{N}  \sum 4\pi iP_{jk}P_{kl}P_{lj}S_{\triangle jkl}\\
  & = &\sigma_{xy}
\end{eqnarray*}
where the summation in the last two lines is over $\vec{r}_{j}$, $\vec{r}_{k}$ and $\vec{r}_{l}$, $S_{\triangle jkl}$ is the signed area of the triangle defined by the three vertex points, and $N$ is the total number of sites. The last equality is based on the fact that the ground state of the original Hamiltonian and $H'$ necessarily belong to the same topological phase and hence we expect their topological quantity $\sigma_{xy}=\sigma_{xy}'$.  This concludes our proof that the Hall conductivity can be expressed in terms of triangular quantum loops consisting of two-point correlators for a gapped system.

\begin{figure}
\includegraphics[scale=0.35]{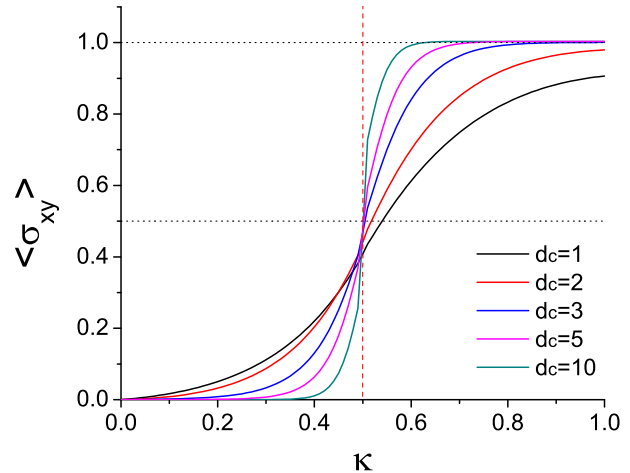}
\caption{The Hall conductivity $\sigma_{xy}$ of Eq. \ref{eq:sqham}
calculated in Eq. \ref{eq:sigmaxy} after summing over triangles
smaller or equal to length scale $d_c=1,2,3,5,10$. To compare, the
topological phase transition is at $\kappa = 0.5$, with $\sigma_{xy}
= 0$ for $\kappa<0.5$ and $\sigma_{xy} = 1$ for
$\kappa>0.5$.}\label{fig:sigmaxy}
\end{figure}

Since $P_{ij}$ decays exponentially as the distance between $i$ and $j$ increases while the areas and number of triangles grow as power-law, the contribution from triangles much larger than the correlation length can be safely neglected. For instance, the Hall conductivity of Eq. \ref{eq:sqham}, $\kappa \in [0,1]$ is shown in Fig. \ref{fig:sigmaxy}, evaluated according to Eq. \ref{eq:sigmaxy} with different cut-off length  $d_c$ for the triangles. The estimation of $\sigma_{xy}$ becomes asymptotically improved as $d_c$ increases, since the inclusion of triangles with longer length scales allows more accurate description when the insulating gap is small and correlation length is long, especially around the transitions.

\section{Interpolating between a Chern insulator and a trivial insulator on the honeycomb lattice}\label{app:honeycomb}
\begin{figure}
\includegraphics[scale=0.35]{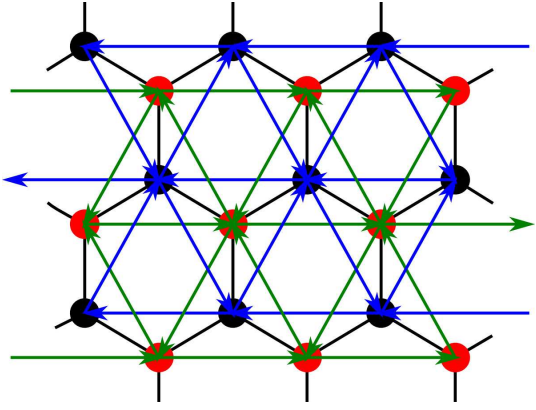}
\caption{The tight-binding model in Eq. \ref{eq:honeycomb} on the honeycomb lattice, where the competition between the imaginary next-nearest neighbor hopping and the staggered on-site potential determines the topological phase of the resulting insulator - Chern insulator at $\kappa > 0.5$ and trivial insulator for $\kappa < 0.5$. The black lines are the nearest neighbor hopping, the green and blue arrows are next-nearest neighbor hopping between the $A$ (red sites) and $B$ (black sites) sublattices, respectively.}\label{fig:honeycomb}
\end{figure}

For completeness, we include in this section the honeycomb lattice model we consider in the main text, which is described by the following Hamiltonian:
\begin{eqnarray}
H&=&\underset{\left\langle ij\right\rangle }{\sum}c_{iB}^{\dagger}c_{jA}+\underset{\left\langle \left\langle ik\right\rangle \right\rangle, s}{\sum}i\kappa\Delta_{ik}c_{is}^{\dagger}c_{ks}+\mbox{h.c.}\nonumber\\
& &+\underset{i}{\sum}3\sqrt{3}(1-\kappa)\Delta\left(c_{iA}^{\dagger}c_{iA}-c_{iB}^{\dagger}c_{iB}\right)
\label{eq:honeycomb}
\end{eqnarray}
where $s=A, B$ labels the two sublattices, the next-nearest neighbor hopping is $i \kappa \Delta$ along the arrows and $-i \kappa \Delta$ against the arrows, see Fig. \ref{fig:honeycomb}. The second line is a staggered on-site potential. We set $\Delta = 0.5$.

The model gives a Haldane's honeycomb Chern insulator model\cite{Haldane1988} for $\kappa = 1$. As $\kappa$ decreases, the system undergoes a topological phase transition to a trivial phase at $\kappa = 0.5$.

\section{Variational Monte Carlo calculations for QLT samples}\label{app:vmc}

In this section, we briefly discuss our algorithm for generating the QLT samples using VMC calculations. Given a many-body wave function, the expectation value of an operator $O$ can be evaluated as:
\begin{eqnarray}
\left\langle O\right\rangle &=& \underset{\alpha\beta}{\sum}\left\langle
\Phi\left|\alpha\left\rangle \cdot\left\langle
\alpha\left|O\right|\beta\right\rangle \cdot\right\langle
\beta\right|\Phi\right\rangle\\ \nonumber
&=&\underset{\alpha}{\sum}\left\langle
\Phi\left|\alpha\left\rangle \cdot\right\langle
\alpha\right|\Phi\right\rangle
\cdot\left[\underset{\beta}{\sum}\left\langle
\alpha\left|O\right|\beta\right\rangle \frac{\left\langle
\beta|\Phi\right\rangle }{\left\langle \alpha|\Phi\right\rangle}\right]
\end{eqnarray}
where $\left|\alpha\right\rangle$ and $\left|\beta\right\rangle$ are real-space wave-function configurations, and $\left\langle \alpha|\Phi\right\rangle$ and $\left\langle \beta|\Phi\right\rangle$ are the respective amplitudes. The first factor in the summation is positive definite and normalized, therefore can serve as the probability density for Monte Carlo sampling, and the second factor is the quantity averaged over the Markov chain.

Similarly,
\begin{eqnarray}
\mbox{tr}\left[P_{jk}P_{kl}P_{lj}\right] &=& \underset {\alpha_{i}}{\sum} \underset{i=1,2,3}{\prod} \left\langle
\Phi\left|\alpha_{i}\left\rangle \cdot\right\langle
\alpha_{i}\right|\Phi\right\rangle \frac{\left\langle
\beta_{i}|\Phi\right\rangle}{\left\langle \alpha_{i}|\Phi\right\rangle}
\nonumber\\&=&\left\langle\tilde{P}_{jk}|_{\alpha_1}\tilde{P}_{kl}|_{\alpha_2}\tilde{P}_{lj}|_{\alpha_3}\right\rangle
\end{eqnarray}
where $\left|\alpha_i\right\rangle$, $i=1,2,3$ are sampled independently and $\left|\beta_{1}\right\rangle=c^\dagger_{j} c_{k}\left|\alpha_{1}\right\rangle$ and so on. In practice, the loop products $\tilde{P}_{jk}|_{\alpha_1}\tilde{P}_{kl}|_{\alpha_2}\tilde{P}_{lj}|_{\alpha_3}$ are sampled over 10 uncorrelated $\left|\alpha_i\right\rangle$ sets to remove the 0-valued inputs and make the learning more efficient.

For the models considered in the main text, $\left\langle
\alpha|\Phi\right\rangle$ takes the form of Slater determinants for
non-interacting Chern insulators, while the FCI states are the third
power of that through parton construction $c=f_1f_2f_3$. $c$ is the
physical fermion operator, and $f_i$, $i=1,2,3$ are operators of
different flavors of parton occupying a $C=1$ Chern insulator each.
The three degenerate ground states can be obtained by threading $\pm
2\pi/3$ fluxes in the parton Chern insulators\cite{FrankTEE}. After
integrating out the partons, the action for the $SU(3)$ gauge field
representing the constraints takes the form of a Chern-Simons term
and a fractional phase, yet breaks down when the parton Chern
insulator has a diminishing gap.

\section{Impact of training models and QLT cut-off on machine learning phases and phase transitions}

In the main text, we have chosen for the training group very typical models in the trivial and topological phases, respectively. Consequently, their correlation length is shorter and more information is distributed over the smaller loops, hence the results' fast convergence in the QLT cut-off $d_c$. Such selection is particularly efficient and effective at recognizing phases, but brings limitations to pinpointing phase transitions as a trade off.

\begin{figure}
\includegraphics[scale=0.35]{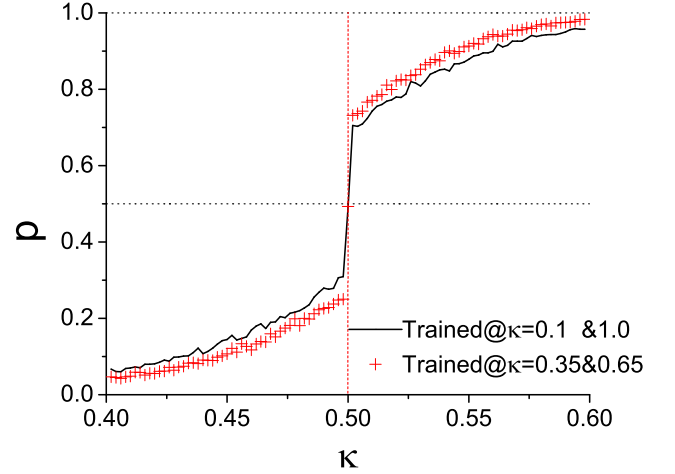}
\caption{Red symbols: the ratio p of `topological' response from the
neural network trained with $\kappa=0.35$ for $y=0$ and
$\kappa=0.65$ for $y=1$. The black curve is the contrast from the
main text with $\kappa=0.1$ and $\kappa=1.0$ in the training group.
Red dashed line marks the expected topological phase transition at
$\kappa = 0.5$. $d_c=3$.}\label{fig:transition}
\end{figure}

To improve the accuracy around $\kappa \sim \kappa_c$, it helps to bring in models with smaller gaps and longer correlations in the training group, as well as an increased $d_c$ for larger loops to distinguish and analyze such information. For example, we repeat the procedures for the square lattice model in the main text, but use training group data from $\kappa=0.35$ and $\kappa=0.65$ for trivial insulator and Chern insulator, respectively. These parameters yield models with smaller gaps and longer correlations therefore better resembles scenarios in the critical regions. As shown in Fig. \ref{fig:transition}, there is a slight improvement near the critical value $\kappa \sim 0.5$ and sharper non-analytical behavior of $p$.

\begin{figure}
\includegraphics[scale=0.35]{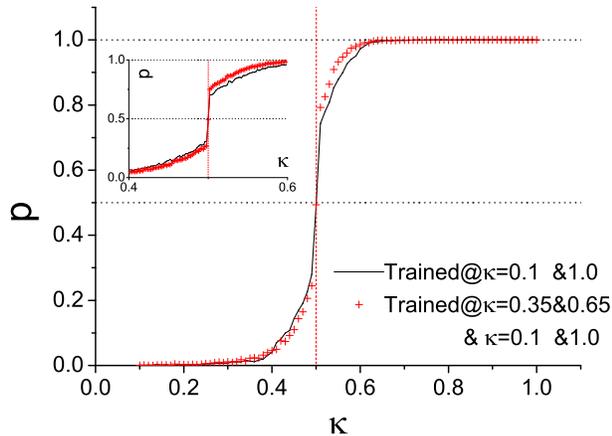}
\caption{Red symbols: the ratio p of `topological' response from the neural network trained with $\kappa=0.1$ and $\kappa=0.35$ for $y=0$ and $\kappa=1.0$ and $\kappa=0.65$ for $y=1$. The black curve is the contrast from the main text with only $\kappa=0.1$ and $\kappa=1.0$ in the training group. Red dashed line marks the expected topological phase transition at $\kappa = 0.5$. $d_c=3$. The inset is an enlargement over the critical region.}\label{fig:finitesize}
\end{figure}

However, training with only $\kappa=0.35$ and $\kappa=0.65$ data does not give the best accuracy on testing groups deep in the trivial or topological phases, since they are less typical and representative for their respective phases. Once again, diversity gives the best overall result, see Fig. \ref{fig:finitesize}, where we include in the training group data from both $\kappa=0.35$ and $\kappa=0.10$ for the trivial insulator, and $\kappa=0.65$ and $\kappa=1.0$ for the Chern insulator.

\end{document}